\documentclass[twocolumn,aps,prl,preprintnumbers]{revtex4-2}
\usepackage{bbm}
\usepackage[latin9]{inputenc}
\usepackage{amsmath}
\usepackage{amssymb}
\usepackage{graphicx}
\usepackage{mathrsfs}
\usepackage{amsfonts}
\usepackage{bm}
\usepackage{amsthm}
\usepackage{xcolor}
\usepackage{txfonts}
\usepackage{float}
\usepackage[colorlinks=true,citecolor=blue,linkcolor=blue,urlcolor=blue,anchorcolor=blue]{hyperref}%
\usepackage{pifont}
\hypersetup{colorlinks=true,citecolor=blue,linkcolor=blue,urlcolor=blue}
\providecommand{\U}[1]{\protect\rule{.1in}{.1in}}
\makeatletter
\@ifundefined{textcolor}{}
{
\definecolor{BLACK}{gray}{0}
\definecolor{WHITE}{gray}{1}
\definecolor{RED}{rgb}{1,0,0}
\definecolor{GREEN}{rgb}{0,1,0}
\definecolor{BLUE}{rgb}{0,0,1}
\definecolor{CYAN}{cmyk}{1,0,0,0}
\definecolor{MAGENTA}{cmyk}{0,1,0,0}
\definecolor{YELLOW}{cmyk}{0,0,1,0}
}

\makeatother
\begin{document}
\title{
Rolling Two-Dimensional Collinear Magnets into Chiral Nanotubes with $p$-Wave Magnetism 
}
\author{Zhejunyu Jin$^{1}$}
\author{Robin R. Neumann$^{1,3}$}
\author{Rodrigo Jaeschke-Ubiergo$^{2}$}
\author{Jairo Sinova$^{2}$}
\author{Alexander Mook$^{1}$}
\email[Contact author: ]{amook@uni-muenster.de}
\affiliation{$^1$Institute of Solid State Theory, University of M\"{u}nster, D-48149 M\"{u}nster, Germany\\
$^2$Institute of Physics, Johannes Gutenberg University Mainz, D-55128 Mainz, Germany\\
$^3$Institute of Physics, Martin Luther University Halle-Wittenberg and Halle-Berlin-Regensburg Cluster of Excellence ``Center for Chiral Electronics'', D-06099 Halle (Saale), Germany}

\begin{abstract}{
$p$-wave magnets are noncollinear compensated magnetic systems that exhibit nonrelativistic antisymmetric spin splitting in momentum space. Their odd-parity spin symmetry enables unconventional spintronic functionalities, including highly efficient charge-to-spin conversion via the Edelstein effect. An outstanding question is whether such magnetic phases can emerge from simple and broadly accessible magnetic building blocks rather than from intrinsically noncollinear magnetic orders. Here, we show that rolling two-dimensional collinear magnets---ferromagnets, antiferromagnets, and altermagnets---into nanotubes generates a rich spin-symmetry landscape controlled by curvature, chirality, and magnetic order. Remarkably, chiral nanotubes hosting radial or tangential coplanar spin textures generically realize $p$-wave magnetism irrespective of the underlying collinear parent phase. The emergent odd-parity spin symmetry manifests itself in both electronic and magnonic spectra through antisymmetric $p$-wave spin splitting. Our results establish magnetic nanotubes as a versatile platform for engineering unconventional $p$-wave magnetism and predict a nonrelativistic Edelstein response that exceeds conventional spin-orbit-driven charge-to-spin conversion by more than an order of magnitude.}
\end{abstract}

\maketitle

\textit{Introduction---}
Spontaneous spin ordering can give rise to momentum-dependent spin splitting even in the absence of spin-orbit coupling. Prominent examples are altermagnets \cite{Smejkal3} and antialtermagnets \cite{Hellenes2023,Jungwirth2025}, which exhibit even- and odd-parity-wave spin splitting, respectively, despite vanishing net magnetization \cite{Smejkal3,Smejkal1,Hayami2020,Ma2022,McClarty2024,Gomonay2024,Ma2021,Duan2025,Osumi2024,Jiang2025,Hellenes2023,Brekke2024,Song2025Electrical}. Their unconventional spin symmetries give rise to nonrelativistic spin-momentum locking, unconventional collective excitations \cite{Smejkal2,Cui2023,Jin2026,Gunnink2026,Yang2026,Neumann2026}, and efficient charge-to-spin conversion \cite{Rafael2021,Bai2023}. In particular, $p$-wave magnets exhibit a nonrelativistic Edelstein effect that can surpass conventional spin-orbit-driven mechanisms \cite{Chakraborty2025}.

The classification of unconventional magnets is naturally formulated in terms of spin symmetries, which allow lattice and spin degrees of freedom to transform independently \cite{Smejkal3,Hellenes2023,Brinkman1996,Litvin1974,Litvin1977}. Beyond the material classification, an important challenge is to identify generic and controllable routes toward realizing unconventional magnetic phases. Existing approaches include strain engineering, symmetry lowering, stacking, moir\'e superstructures, electric-field control, and surface or interface effects \cite{Chakraborty2024,Zhou2025,Karetta2025,Belashchenko2025,Khodas2025,Shan2025,Liu2025,Wang2025,Zeng2024,Lange2026,Sasioglu2026,Leeb2026}. While these strategies successfully manipulate crystal and spin symmetries, they typically preserve collinear spin order and therefore mainly enable transitions within the manifold of collinear unconventional magnets. Odd-parity-wave magnets may also be engineered through Floquet driving \cite{Li2026Floquet,Huang2026Light,Zhu2025Floquet,Liu2026Light} or loop-current order \cite{Leeb2026Collinear}; however, these routes rely on nonequilibrium or orbital-current physics. In static spin-order-driven systems, odd-parity-wave magnetism generally requires noncollinear magnetic textures \cite{Hellenes2023}, whose controlled realization remains challenging.
        
In parallel, curved magnetic systems have emerged as a versatile platform in which geometry and topology reshape magnetic interactions \cite{Kravchuk2018,Yershov2025,Bittencourt2022,Yershov2022,Yang2022,Kravchuk2016,Yan2025,Gobel2025}. Curvature can induce effective anisotropies and chiral couplings that are absent in flat systems, thereby stabilizing noncollinear magnetic textures. In particular, magnetic nanotubes provide a highly tunable realization of curved magnetism \cite{Landeros2009,Yan2012,Otlora2017,KorberMode2022,Gallardo2022,KorberCurvilinear2022}, where the interplay between geometry and spin order may lead to qualitatively new magnetic states.
   
\begin{figure}
    \centering
    \includegraphics[width=\columnwidth]{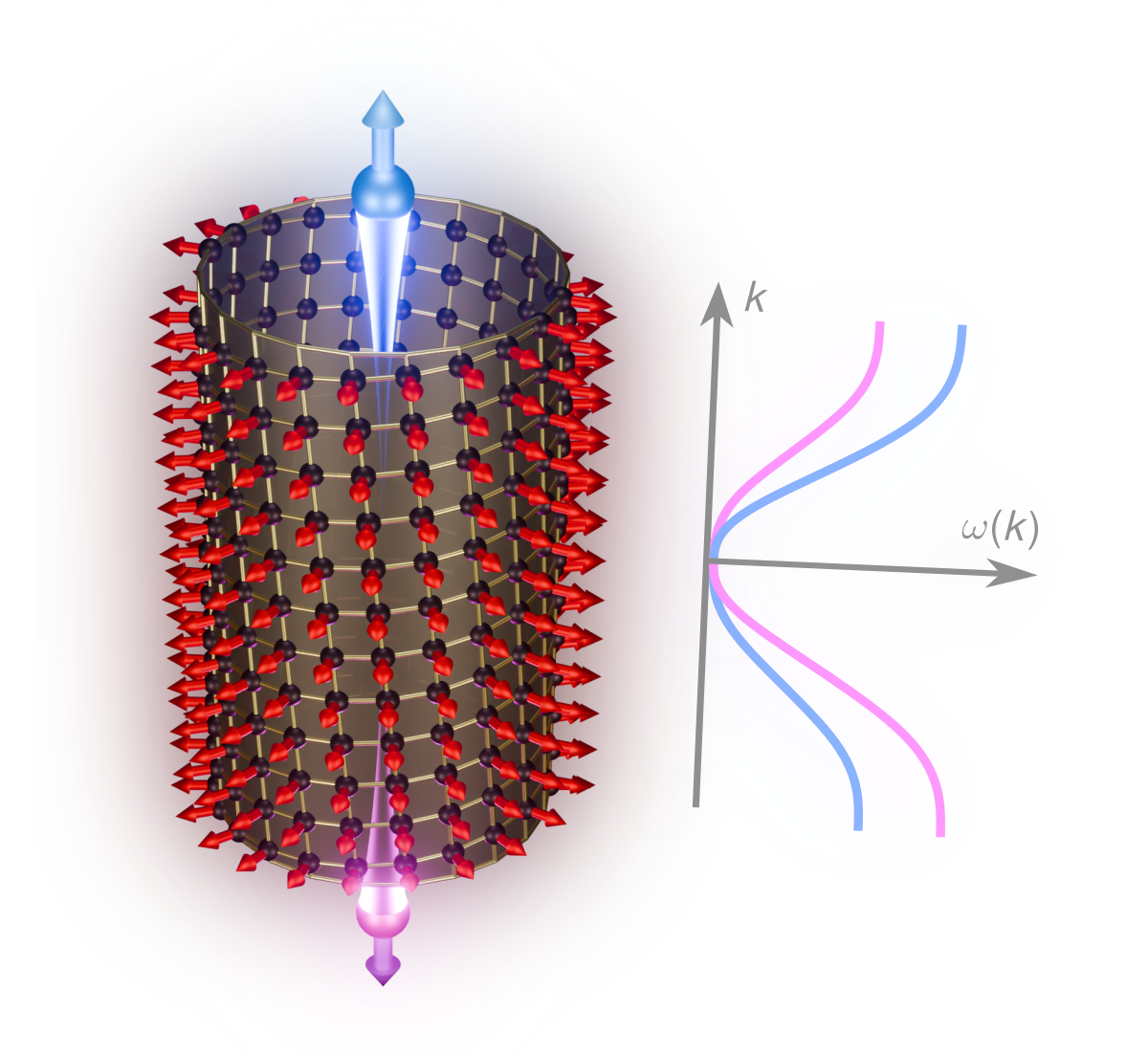}
    \caption{$p$-wave magnetism in chiral nanotubes with coplanar radial spin configuration (red arrows). The coplanar spin symmetry $[C_{2,z'}\mathcal T\|\mathcal T]$, together with the screw constraint of the chiral tube, produces a time-reversal-symmetric band structure, while broken inversion symmetry allows antisymmetric spin splitting. The itinerant spin-up and spin-down electrons (cyan and magenta spheres) move in opposite direction due to nonrelativistic spin-momentum locking, which manifests in an antisymmetric band structure: $\omega_{\uparrow}(k) = \omega_{\downarrow}(-k)$. This  $p$-wave spin splitting emerges in chiral nanotubes built from ferromagnets, antiferromagnets, or altermagnets.}
    \label{fig1}
\end{figure}

\begin{figure*}[htbp] 
  \centering
  \includegraphics[width=1\textwidth]{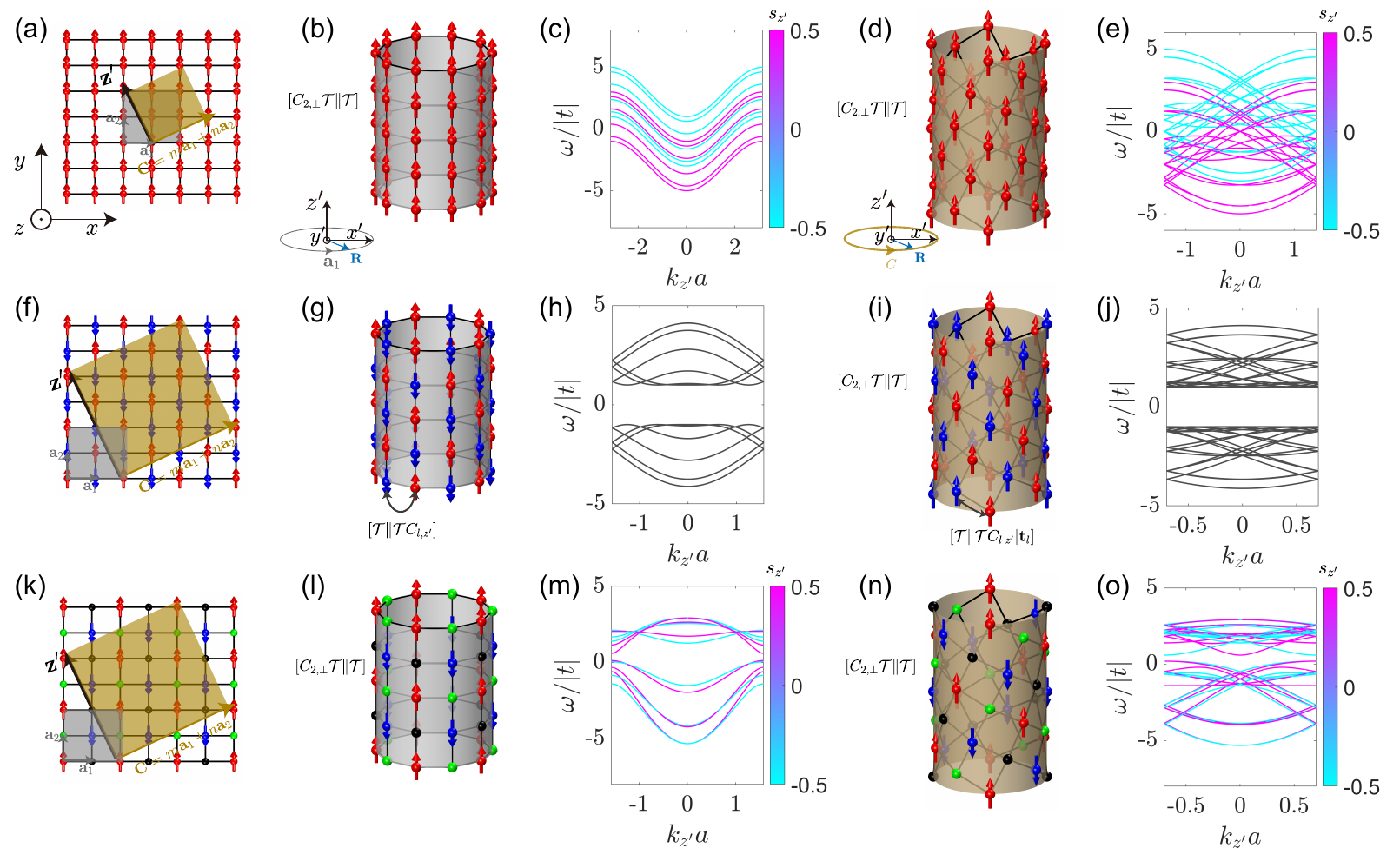}\\
  \caption{Parent magnetic orders (a,f,k), corresponding collinear magnetic textures (b,d,g,i,l,n), and electronic band structures (c,e,h,j,m,o) of achiral (silver-gray) and chiral (gold) nanotubes derived from planar ferromagnets (b-e), antiferromagnets (g-j), and altermagnets (l-o). Black and green spheres denote inequivalent nonmagnetic atoms. Gray curves indicate spin-degenerate bands, while colored curves represent spin-split bands. The spin textures are aligned along the $y$ direction in the planar structures and along the $z'$ direction in the nanotubes. For ferromagnetic and antiferromagnetic nanotubes, we use nearest-neighbor hopping with $t_{i,j}=J_{sd}=t$ and $\mu=0$. For altermagnetic nanotubes, next-nearest-neighbor hopping is included as follows: along the $(1,1)$ direction, the hopping amplitude is $0.4t$ ($0.2t$) on the red (blue) sublattice, whereas along the $(1,-1)$ direction it is $0.2t$ ($0.4t$) on the red (blue) sublattice. The achiral tube has circumference $10a$, and the chiral tube is specified by $(m,n)=(8,4)$, where $a$ denotes the nearest-neighbor bond length. Spin symmetries responsible for collinear spin textures in momentum space or Kramers-degenerate spectra are indicated. The symmetry $[C_{2,\perp} \mathcal{T}\| \mathcal{T}]$ imposes an inversion-symmetric band structure in the nanotubes. Kramers degeneracy is lifted by the finite magnetization in (b-e), whereas it is protected by combined inversion symmetry and $[\mathcal{T}\| \mathcal{T}|C_{l,z'}]$ in (g,h), and by combined inversion symmetry and $[C_{l,\perp}\mathcal{T}\| \mathcal{T}|{\bf t}_l]$ in (i,j). Notably, in altermagnets there is no spatial symmetry connecting opposite spins, and Kramers degeneracy is therefore lifted in (l-o).}\label{fig2}
\end{figure*}

In this work, we establish a direct connection between curved magnetism and unconventional spin-split phases. We show that rolling a two-dimensional collinear magnet---including ferromagnets, antiferromagnets, and altermagnets---into a nanotube geometry can generically transform the magnetic order into an odd-parity-wave phase. Crucially, in the presence of easy-axis anisotropy along the surface normal, curvature naturally stabilizes a radial spin texture in which the local magnetic moments align perpendicular to the tube surface. This radial noncollinear state universally realizes $p$-wave magnetism in chiral nanotubes, independent of the specific collinear order in the underlying planar system. As illustrated in Fig.~\ref{fig1}, curvature therefore does not merely perturb the magnetic state or lower crystal symmetries, but instead provides a geometric mechanism that fundamentally reorganizes the spin texture and converts collinear magnets into unconventional odd-parity-wave phases. The resulting nonrelativistic spin-momentum locking generates a pronounced Edelstein effect, enabling efficient electrical generation of spin polarization. For experimentally realistic nanotube diameters, we predict charge-to-spin conversion efficiencies that exceed conventional spin-orbit-driven Edelstein responses by more than an order of magnitude. Our results establish curved geometries as a promising route toward engineering unconventional magnetism and identify magnetic nanotubes as a practical platform for realizing $p$-wave magnetic symmetry and highly efficient spin-charge interconversion.

\textit{Collinear magnetic nanotubes---} We first consider magnetic nanotubes in the absence of magnetic anisotropy. In this limit, all spins align parallel due to isotropic Heisenberg exchange, forming a collinear configuration that we take along the $z'$ axis without loss of generality, as shown in Fig.~\ref{fig2}. The resulting electronic spin polarization is fully along the magnetic order direction, $s_{z'}$, while the transverse components $s_{x'}$ and $s_{y'}$ vanish in momentum space. In the nonrelativistic limit, the system preserves the combined spin symmetry $[C_{2,\perp}{\mathcal T}\|{\mathcal T}]$. [The operations acting in spin space (left of $\|$) and those acting in real space (right of $\|$) can be generally distinct in the absence of spin-orbit coupling \cite{Brinkman1996,Litvin1974,Litvin1977}.]
Here, $\mathcal T$ denotes time reversal symmetry, which inverts the spin and crystal momentum, while $C_{2,\perp}$ denotes a $180^\circ$ spin rotation about an axis perpendicular to the ordered moment. Their combination enforces an even-parity spin texture, $\omega(s_{z'},{\bf k})=\omega(s_{z'},-{\bf k})$. Consequently, collinear magnetic nanotubes allow only two possibilities: either $s$-wave spin splitting or spin-degenerate bands. $p$-wave spin splitting is therefore excluded in the collinear regime. 

Inspired by carbon nanotubes \cite{Ajayan1997,Laird2015,Zhang2017}, we consider two rolling geometries: (i) achiral nanotubes formed along a lattice vector ${\bf a}_1$ or ${\bf a}_2$, and (ii) chiral nanotubes formed along a vector ${\bf C}=m{\bf a}_1+n{\bf a}_2$ with $m\neq n$ [see Figs.~\ref{fig2}(a), (f), and (k)].
Below we perform a symmetry analysis and tight-binding calculations [see Eq.~\eqref{Eq1} in the End Matter].

We first focus on ferromagnetic nanotubes. Both achiral and chiral geometries break time-reversal symmetry and therefore possess a finite net magnetization [Figs.~\ref{fig2}(b) and (d)] and $s$-wave spin splitting [Figs.~\ref{fig2}(c) and (e)].

For antiferromagnetic nanotubes, the underlying symmetry operations differ between achiral and chiral geometries. Nevertheless, both configurations preserve effective symmetry constraints that enforce spin degeneracy in the band structure. Specifically, achiral tubes preserve $[\mathcal T||\mathcal T C_{l,z'}]$, whereas chiral tubes preserve $[\mathcal T||\mathcal T C_{l,z'}|{\bf t}_l]$ [Figs.~\ref{fig2}(g) and (i)]. Since the effective model contains only one spin-polarization component and one momentum component, both $[\mathcal T||\mathcal T C_{l,z'}]$ and $[\mathcal T||\mathcal T C_{l,z'}|{\bf t}_l]$ act as effective time-reversal symmetries in momentum space. (The notation before and after the vertical bar denotes the point-group and translational parts of the real-space operation, respectively.) Combined with the effective inversion symmetry associated with $[C_{2,\perp}\mathcal T||\mathcal T]$, these operations enforce spin degeneracy throughout the Brillouin zone in both achiral and chiral nanotubes [Figs.~\ref{fig2}(h) and (j)]. Here, $C_{l,z'}$ denotes a rotation about the nanotube axis $z'$ by $2\pi/l$, where $l$ is fixed by the discrete rotational symmetry of the nanotube cross section. In chiral nanotubes, this rotation is accompanied by a translation ${\bf t}_l$ along the tube axis, forming a screw operation. 

In this sense, collinear ferromagnetic and antiferromagnetic nanotubes can be viewed as effective one-dimensional descendants of their planar parent systems: ferromagnetic tubes retain conventional $s$-wave spin splitting, whereas antiferromagnetic tubes remain spin degenerate.
A similar viewpoint also applies to altermagnetic nanotubes. However, because altermagnetic spin splitting is strongly constrained by the momentum dependence of the parent phase, its one-dimensional projection depends sensitively on the rolling direction. For generic rolling directions, the spin degeneracy is lifted even at the $\Gamma$ point, giving rise to an effective $s$-wave magnetic phase [Figs.~\ref{fig2}(m) and (o)]. By contrast, when the tube axis is aligned with a nodal direction of the parent altermagnet, corresponding to the special chiral angle $\theta_s=\arctan(n/m)=\pi/4$, the projected nanotube spectrum becomes spin degenerate. Our explicit band calculations confirm this prediction, revealing spin-degenerate bands protected by $[\mathcal{T}||\mathcal{T}{\mathcal M}_{z'}]$ symmetry in the collinear nanotube, where ${\mathcal M}_{z'}$ denotes the mirror symmetry with respect to the plane perpendicular to the $z'$ axis [see Sec.~I of the Supplemental Material (SM) \cite{SM}].

The remaining symmetry operations of collinear nanotubes and their corresponding band-structure characteristics are summarized in Table \ref{table1} (see End Matter).
\begin{figure*}[htbp] 
  \centering
  \includegraphics[width=1\textwidth]{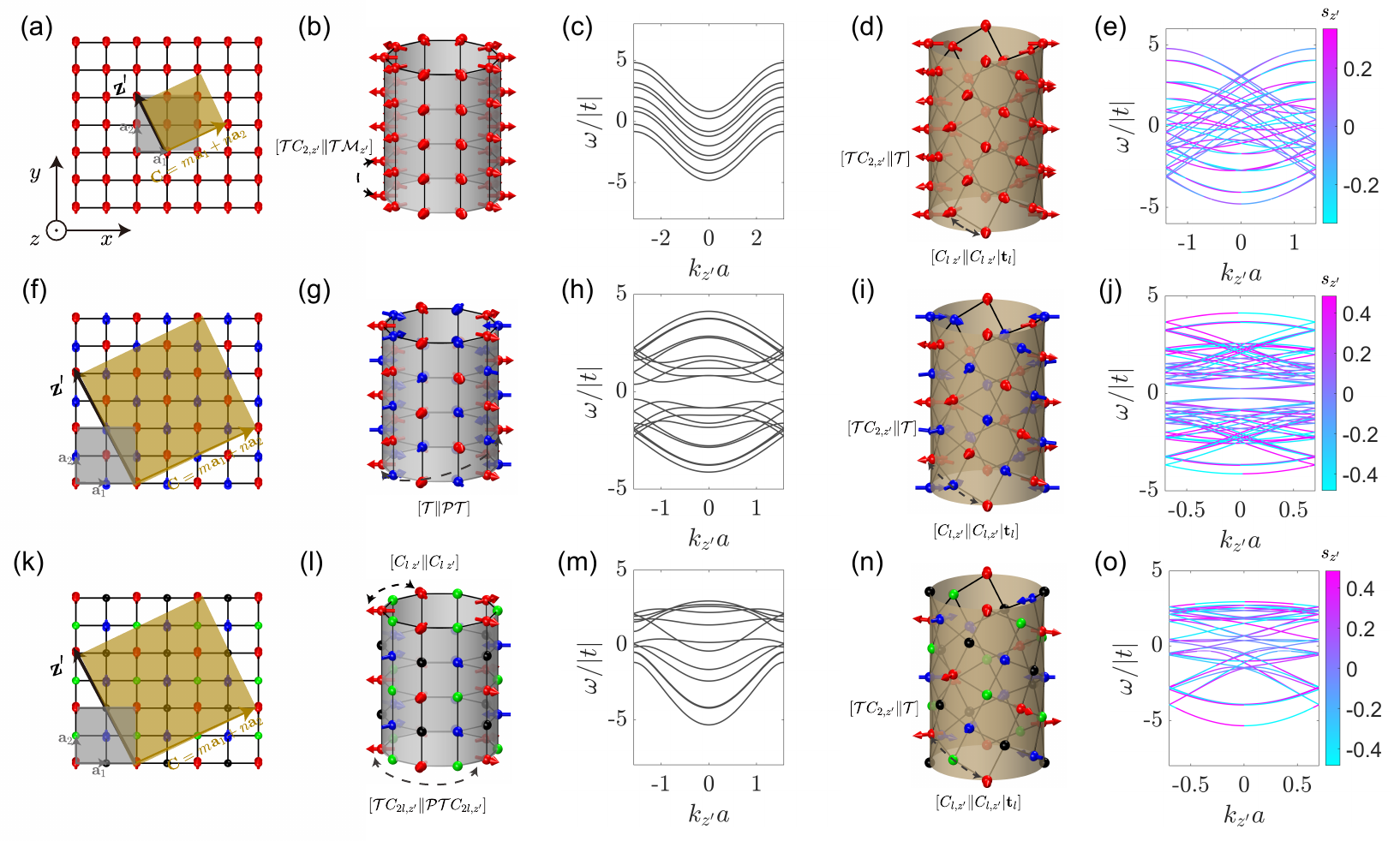}\\
  \caption{Parent magnetic orders (a,f,k), corresponding noncollinear magnetic textures (b,d,g,i,l,n), and electronic band structures (c,e,h,j,m,o) of achiral (silver-gray) and chiral (gold) nanotubes derived from planar ferromagnets (b-e), antiferromagnets (g-j), and altermagnets (l-o).  Black and green spheres denote inequivalent nonmagnetic atoms. Gray curves indicate spin-degenerate bands, while colored curves represent spin-split bands. The spin textures are aligned along the $z$ direction in the planar structures and along the radial direction in the nanotubes. For ferromagnetic and antiferromagnetic nanotubes, we use nearest-neighbor hopping with $t_{i,j}=J_{sd}=t$ and $\mu=0$. For altermagnetic nanotubes, next-nearest-neighbor hopping is included as follows: along the $(1,1)$ direction, the hopping amplitude is $0.4t$ ($0.2t$) on the red (blue) sublattice, whereas along the $(1,-1)$ direction it is $0.2t$ ($0.4t$) on the red (blue) sublattice. The achiral tube has circumference $10a$, and the chiral tube is specified by $(m,n)=(8,4)$, where $a$ denotes the nearest-neighbor bond length. The collinear spin texture in momentum space is protected by rotational symmetry in achiral tubes and by screw symmetry in chiral tubes. In the coplanar tubes, the symmetry $[C_{2,z'} \mathcal{T}\|\mathcal{T}]$ relates $(s_{z'},k)$ to $(-s_{z'},-k)$, while screw symmetry suppresses transverse spin components. Together these constraints produce a time-reversal-symmetric band structure. Kramers degeneracy is protected by combined symmetry constraints in the achiral tubes (b,c), (g,h), and (l,m). In chiral tubes (d,e,i,j,n,o), inversion symmetry is broken, and the time-reversal-symmetric band structure therefore exhibits odd-parity spin splitting.}\label{fig3}
\end{figure*} 

\textit{Noncollinear magnetic nanotubes---}The collinear nanotubes discussed above either retain even-parity spin splitting or remain spin degenerate, but they do not realize $p$-wave spin splitting. We now turn to the case where perpendicular magnetic anisotropy stabilizes noncollinear magnetic textures, placing the nanotube in a qualitatively different symmetry setting. 

We consider planar magnets with out-of-plane easy-axis anisotropy. Upon rolling them into a nanotube, this anisotropy transforms into a radial magnetic anisotropy, which we assume to be strong enough to favor spin alignment along the local radial direction (which is the case above a characteristic tube radius \cite{Edstr2022, Caha2024}). Consequently, the system develops a coplanar magnetic configuration. Despite the noncollinearity, the symmetry $[C_{2,z'}{\mathcal T}||{\mathcal T}]$ remains preserved in both achiral and chiral nanotubes. This symmetry constrains the spin component along the rotation axis to be odd in ${\bf k}$, while the perpendicular components are even. As a result, it enforces the relation $\omega(s_{z'},{\bf k})=\omega(-s_{z'},-{\bf k})$. When the transverse spin components are additionally suppressed, $s_{x'}=s_{y'}=0$, the resulting spin polarization in momentum space becomes collinear and antisymmetric.

Achiral nanotubes derived from planar ferromagnets additionally preserve the $[E||\mathcal M_{z'}]$ symmetry, which acts as an effective inversion symmetry in the one-dimensional momentum space. The transverse spin polarizations are suppressed by the rotational symmetry, leaving only the $s_{z'}$ component \cite{SM}. Together with the previously discussed $[C_{2,z'}\mathcal T||\mathcal T]$ constraint, this leads to spin-degenerate bands in momentum space [Figs.~\ref{fig3}(b) and (c)]. Consequently, the rolled structure becomes effectively antiferromagnetic. In contrast, chiral nanotubes preserve a screw symmetry $[C_{l,z'}\| C_{l,z'}|{\bf t}_l]$, which suppresses all transverse spin polarizations \cite{SM}, and break the $[E||{\mathcal M}_{z'}]$ symmetry [Fig.~\ref{fig3}(d)]. In this case, the combined constraints from the coplanar spin symmetry and the screw symmetry enforce a time-reversal-symmetric band structure. Since inversion symmetry is broken by the chiral lattice geometry and the noncollinear magnetic texture, the system exhibits odd-parity spin splitting, $\omega(s_{z'},{\bf k})=\omega(-s_{z'},-{\bf k})$ [Fig.~\ref{fig3}(e)].

For antiferromagnetic planar parent magnets, the achiral nanotube preserves the combined $[{\mathcal T}\|{\mathcal P}{\mathcal T}]$ symmetry, resulting in spin-degenerate bands in momentum space [Figs.~\ref{fig3}(g) and (h)]. For the chiral nanotube, the system breaks $[{\mathcal T}\|{\mathcal P}{\mathcal T}]$, but possesses a screw symmetry $[C_{l,z'}\| C_{l,z'}|{\bf t}_{l}]$, which suppresses all transverse spin polarizations [Fig.~\ref{fig3}(i)]. As a result, the chiral nanotube exhibits $p$-wave spin splitting [Fig.~\ref{fig3}(j)].

Finally, we consider a two-dimensional $d$-wave altermagnet as the planar parent compound. For the achiral nanotube, the system preserves the symmetries
$[C_{l,z'}|| C_{l,z'}]$ and $[{\mathcal T}C_{2l,z'} ||{\mathcal P}{\mathcal T}C_{2l,z'}]$ [Fig.~\ref{fig3}(l)]. For an odd number of spins along the circumference, the operation $[{\mathcal T}C_{2l,z'} ||{\mathcal P}{\mathcal T}C_{2l,z'}]$ requires the additional rotation $[C_{2l,z'} || C_{2l,z'}]$, whereas this extra rotation is not needed for an even number of spins. The former symmetry suppresses transverse spin polarization, while the latter acts as an effective $[{\mathcal T}\|{\mathcal P}{\mathcal T}]$ symmetry, since the system involves only a single spin component and a single momentum direction. The combined effect reconstructs the magnetic state into an effective antiferromagnet with spin-degenerate bands in momentum space [Fig.~\ref{fig3}(m)]. For the corresponding chiral nanotube [see Fig.~\ref{fig3}(n)], the symmetry structure is analogous to that of the ferromagnetic and antiferromagnetic case, resulting in a collinear $p$-wave spin texture in momentum space [Fig.~\ref{fig3}(o)]. Moreover, when the tube axis is aligned parallel to a nodal line of the parent altermagnet, the nanotube belongs to the achiral class, and spin-degenerate bands are protected by $[E||{\mathcal P}]$ and $[C_{2,z'} \mathcal{T}\| \mathcal{T}]$ symmetry \cite{SM}.

We summarize the symmetries in coplanar magnetic tubes in Table~\ref{table2}. 

\begin{figure}[htbp] 
  \centering
  \includegraphics[width=0.96\columnwidth]{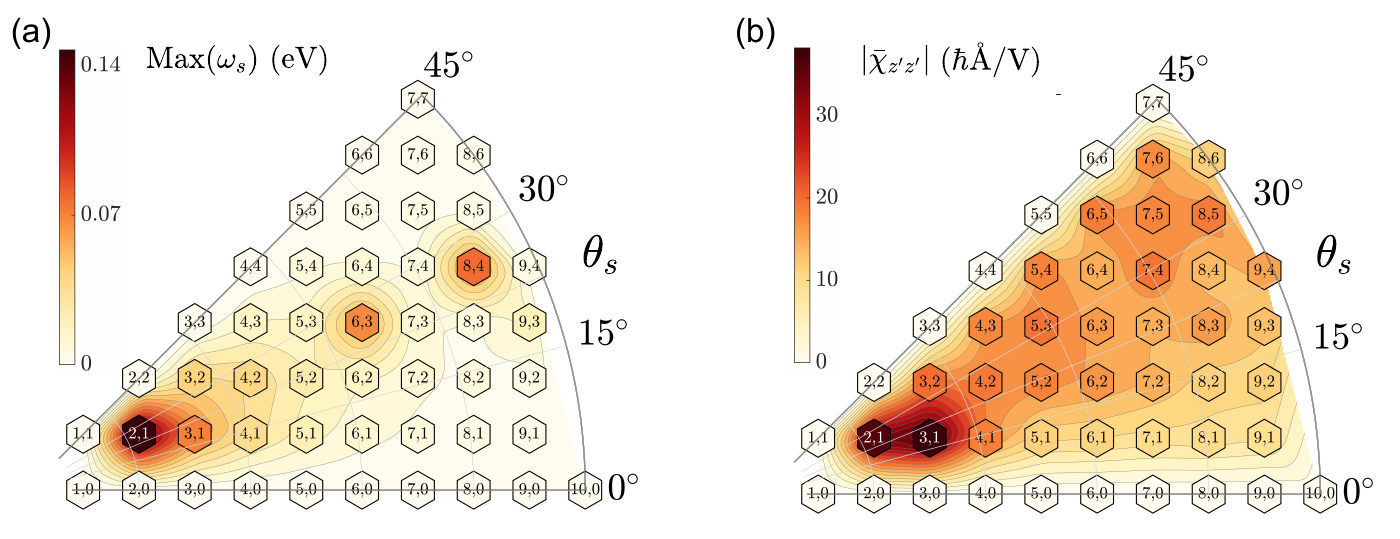}\\
  \caption{(a) Maximum $p$-wave spin splitting $\omega_s$ and average Edelstein response $\overline{|\chi_{z'z'}|}$
  as a function of ($m$, $n$) in  noncollinear tubes derived from planar ferromagnets. Parameters read: $t_{i,j}=J_{sd}=1$ eV and $T=0$ K.}\label{fig4}
\end{figure}

\textit{Geometric tunability of $p$-wave spin splitting---}
The odd-parity spin splitting originates from the coplanar spin texture along the helical pitch direction and is therefore controlled by the winding indices $(m,n)$. As the chiral angle approaches the achiral limits $0$ or $\pi/4$, the bands become spin degenerate; accordingly, the $p$-wave splitting is suppressed near these high-symmetry limits and maximized at intermediate chiral angles with $m=2n$ [Fig.~\ref{fig4}(a)]. We quantify the splitting by the maximum energy separation between the two spin-polarized branches of the lowest band over the Brillouin zone, which provides a clean measure unaffected by the many subbands near the Fermi level. Sizable splittings near the Fermi level are discussed in the SM \cite{SM}.

The spin splitting also decreases with increasing tube radius [Fig.~\ref{fig4}(a)]. In the large-radius limit, the nanotube approaches a planar geometry with periodic boundary conditions along the chiral vector, and the spectrum reduces to folded copies of the planar band structure, causing the odd-parity splitting to vanish. Small oscillations may occur due to subband folding and band-character switching, but they do not affect the overall decreasing trend with tube size \cite{SM}.

Overall, the largest $p$-wave spin splitting occurs for tubes with small radius and intermediate chiral angle. Finally, tubes with opposite handedness, such as $(-m,n)$ or $(m,-n)$, exhibit spin splitting with the opposite sign to that of the $(m,n)$ tube, as required by inversion \cite{SM}.

\textit{Edelstein effect in $p$-wave chiral nanotubes---}
We next consider the electric-field response in $p$-wave magnetic nanotubes. In equilibrium, the net spin polarization vanishes over the Brillouin zone due to the odd-parity spin texture in momentum space. When an electric field is applied, the system is driven out of equilibrium, leading to an asymmetric redistribution of electrons in momentum space. This imbalance generates a finite magnetization, known as the Edelstein effect \cite{Chakraborty2025}.
Within linear response theory, the induced magnetization is given by $\langle M_{\mu}\rangle = \chi_{\mu\nu} E_{\nu}$, with the response tensor
\begin{equation}\label{Eq2}
\begin{aligned}
\chi_{\mu\nu} = \sum_n\frac{e\tau}{\hbar} g\mu_B \int \frac{dk}{2\pi/a}\, s_{n,\mu}(k)\, v_{n,\nu}(k) \frac{\partial f^{(0)}_n}{\partial \omega_n},
\end{aligned}
\end{equation}
where $g$ is the Land\'{e} g-factor, $\mu_B$ the Bohr magneton, $e$ the elementary charge, ${\bf v}$ the group velocity, $\tau$ the relaxation time, and $f^{(0)}_n=1/\{\exp[(\omega_n-\mu)/k_B T]+1\}$ the Fermi distribution.

The Edelstein response is more intricate than the spin splitting, since it depends on the chemical potential, the number of Fermi-crossing bands, and their group velocities, in addition to the spin polarization. To characterize the overall geometric trend, we use the chemical-potential-averaged response $\overline{|\chi_{z'z'}|}$, obtained by averaging $|\chi_{z'z'}(\mu)|$ over the full spectrum.

As shown in Fig.~\ref{fig4}(b), the largest response appears for tubes with small radius and intermediate chiral angle. This is consistent with the spin-splitting analysis: small-radius tubes retain stronger noncollinearity and spin polarization, while intermediate chiral angles avoid the two achiral limits where the odd-parity response is symmetry suppressed. Using a representative metallic relaxation time $\tau=50$ fs and lattice constant $a=4$ \AA, comparable to two-dimensional magnetic metals such as Fe$_3$GeTe$_2$ \cite{Badrtdinov2023}, the effective Edelstein response integrated over one primitive tube unit cell reaches $\sim 40~\hbar{\rm \AA}/\text{V}$, where V is volt. Even for relatively large tubes, it remains of order $\sim 20~\hbar{\rm \AA}/\text{V}$. These values exceed the spin-orbit-induced Edelstein response reported for LuFeO$_3$\cite{Hernandez2024} and are comparable to those predicted in other $p$-wave magnets \cite{Chakraborty2025}, suggesting that the effect should be experimentally observable.

We emphasize that the spin splitting and the Edelstein response are purely nonrelativistic. For a discussion of spin-orbit coupling, see End Matter and Sec.~V of the SM \cite{SM}.

\textit{Discussion---}Magnetic nanotubes with coplanar order have recently been realized in CrI$_3$ \cite{Caha2024}, an honeycomb-lattice van der Waals magnet. 
The symmetry-based results obtained for the square lattice carry over directly to the honeycomb lattice (see Sec.~VI of the SM \cite{SM}). Since CrI$_3$ is an insulator, its elementary magnetic excitations are magnons. Because the symmetry arguments underlying $p$-wave magnetism apply equally to magnons \cite{Neumann2026,Kravchuk2026}, the magnon magnetic moment likewise exhibits $p$-wave symmetry in coplanar chiral nanotubes (see Sec.~VII of the SM \cite{SM}).

\textit{Conclusion---}We have investigated the emergence of unconventional spin symmetries in magnetic nanotubes obtained by rolling up collinear magnets. While nanotubes retaining collinear order largely inherit the band properties of their planar parent systems-with the notable exception of altermagnets, where the dominant beyond-$s$-wave spin splitting is converted into an emergent $s$-wave form-chiral nanotubes exhibit a qualitatively different regime driven by curvature-induced coplanar spin textures. These systems realize odd-parity ($p$-wave) magnetism, in which both electronic and magnonic excitations display characteristic $p$-wave spin splitting. This curvature-induced spin symmetry leads to a pronounced nonequilibrium spin accumulation via the Edelstein effect. For experimentally realistic parameters, we find that the resulting charge-to-spin conversion efficiency exceeds that of spin-orbit-coupling-driven mechanisms by more than an order of magnitude. Our results establish a direct link between curvilinear magnetism and unconventional spin-split phases, and identify magnetic nanotubes as a versatile platform for engineering odd-parity magnetism and efficient spin-charge interconversion in low-dimensional magnetic systems.

\begin{acknowledgments}
\textit{Note added---} During the final stages of manuscript preparation, we were informed of recent related work \cite{Ersoy2026}, which investigates altermagnets rolled into chiral nanotubes with collinear magnetic order. The reported dependence of the spin splitting on the chiral angle is in agreement with the behavior found in our study. Their work focuses on collinear altermagnetic nanotubes and does not address the odd-parity spin splitting stabilized by noncollinear magnetic order in chiral nanotubes.

\textit{Acknowledgments---}A.~M.~is indebted to Lukas K\"{o}rber for useful discussions on curvilinear magnetism.
This work was funded by the German Research Foundation (DFG) through TRR 173-268565370 (Projects No.~A03 and B13), TRR 288-422213477 (Projects No.~A09 and B05), and Project No.~504261060 (Emmy Noether Programme). We acknowledge support by the Dynamics and Topology Center (TopDyn) funded by the State of Rhineland-Palatinate. R.~R.~N acknowledges funding by DFG as part of the German Excellence Strategy-EXC3112/1-533767171 (Center for Chiral Electronics). Z.~J. acknowledges financial support from the Alexander von Humboldt postdoctoral fellowship and the National Natural Science Foundation of China (Grant No. 12404125).
\end{acknowledgments}

\setcounter{equation}{0}  
\renewcommand{\theequation}{A\arabic{equation}}  
\setcounter{figure}{0}  
\renewcommand{\thefigure}{A\arabic{figure}}  
\onecolumngrid 
\section*{End Matter}
\twocolumngrid
\textit{Appendix A:Tight-binding model---}
To verify our symmetry analysis, we employ a tight-binding model with isotropic $s$-$d$ exchange interaction,
\begin{equation}\label{Eq1}
\begin{aligned}
H_e=-\sum_{i,j,\sigma} t_{i,j}c_{i,\sigma}^{\dagger}c_{j,\sigma}-\mu\sum_{i,\sigma} c_{i,\sigma}^{\dagger}c_{i,\sigma}-J_{sd}\sum_{i,\sigma,\sigma'}{\bf S}_i\cdot c_{i,\sigma}^{\dagger} {\bm\sigma}_{\sigma\sigma'}c_{i,\sigma'},
\end{aligned}
\end{equation}
where $c_{i,\sigma}^{(\dagger)}$ is the electron (creation) annihilation operator for an electron with spin $\sigma$ at site $i$. The parameter $t_{i,j}$ corresponds to hopping between different sites, $\mu$ is the chemical potential, and $J_{sd}$ is the isotropic $sd$ coupling between a localized spin and itinerant electrons. For ferromagnetic and antiferromagnetic tube, we consider isotropic nearest neighbour (NN) hopping. For altermagnetic tube, an extra anisotropic next nearest neighbour (NNN) hopping is included in tight-binding model. 

\textit{Appendix B: Symmetries and resulting band properties of magnetic nanotubes---}
Rolling a planar crystal into a nanotube does not eliminate all real-space symmetries of the parent structure; rather, several of them are transformed into symmetry operations of the cylindrical geometry. In achiral nanotubes, translations of the planar lattice are mapped onto rotations around the tube axis and, depending on the chosen direction, to mirror-related operations such as $[E||{\mathcal M}_{z'}]$. In chiral nanotubes, the corresponding planar translations are instead mapped onto screw operations combining rotation and translation along the tube axis. In addition, inversion symmetry of the parent planar structure is represented in the nanotube geometry by a twofold rotation about an axis perpendicular to the tube axis. To make these correspondences explicit, we derive below how the relevant twofold rotations in the cylindrical geometry act on the original planar coordinates.
\begin{table*}[htbp]
\centering
\caption{Critical symmetries and resulting band properties of collinear magnetic nanotubes}
\label{table1}
\setlength{\tabcolsep}{8pt}
\begin{tabular}{c c c c}
\hline\hline
Parent magnetic order & Geometry & Symmetry & Band property \\
\hline\hline

FM & Achiral & $[C_{2,\bot}{\mathcal T}\|{\mathcal T}]$, $[E\|{\mathcal M}_{z'}]$ $[E\|C_{l,z'}]$, $[E\|C_{2,\bot}]$, $[E||\mathcal{P}]$ & $\omega({\bf s},{\bf k})=\omega({\bf s},-{\bf k})$, $\omega({\bf s},{\bf k})\neq\omega({\bf -s},{\bf k})$ \\
   & Chiral  & $[C_{2,\bot}{\mathcal T}\|{\mathcal T}]$, $[E\|C_{2,\bot}]$& $\omega({\bf s},{\bf k})=\omega({\bf s},-{\bf k})$, $\omega({\bf s},{\bf k})\neq\omega({\bf -s},{\bf k})$ \\

\hline\hline

AFM & Achiral & $[C_{2,\bot}{\mathcal T}\|{\mathcal T}]$, $[{\mathcal T}\|C_{2,\bot}{\mathcal T}]$, $[{\mathcal T}\|{\mathcal T}C_{l,z'}]$, $[{\mathcal T}\| {\mathcal T}|t_{z'}]$, $[{\mathcal T}\|{\mathcal P}{\mathcal T}]$ & $\omega({\bf s},{\bf k})=\omega({\bf s},-{\bf k})$, $\omega({\bf s},{\bf k})=\omega({\bf -s},{\bf k})$ \\
    & Chiral  & $[C_{2,\bot}{\mathcal T}\|{\mathcal T}]$, $[{\mathcal T}\|{\mathcal T}C_{l,z'}|{\bf t}_{l}]$, $[{\mathcal T}\|C_{2,\bot}{\mathcal T}]$ & $\omega({\bf s},{\bf k})=\omega({\bf s},-{\bf k})$, $\omega({\bf s},{\bf k})=\omega({\bf -s},{\bf k})$ \\

\hline\hline

AM & Achiral & $[C_{2,\bot}{\mathcal T}\|{\mathcal T}]$, $[C_{l,z'}||\mathcal{P}C_{l,z'}]$ & $\omega({\bf s},{\bf k})=\omega({\bf s},-{\bf k})$, $\omega({\bf s},{\bf k})\neq\omega({\bf -s},{\bf k})$ \\
   & Chiral  & $[C_{2,\bot}{\mathcal T}\|{\mathcal T}]$, $[E\|C_{2,\bot}]$ & $\omega({\bf s},{\bf k})=\omega({\bf s},-{\bf k})$, $\omega({\bf s},{\bf k})\neq\omega({\bf -s},{\bf k})$ \\

\hline\hline
\end{tabular}
\end{table*}

We start from a planar coordinate system $(x,y)$ and roll the sheet into a cylinder of radius $R$. The tube axis is chosen along the planar $y$ direction, such that
\begin{equation}\label{Eq2}
z'=y.
\end{equation}
The planar $x$ coordinate is mapped to the azimuthal angle $\phi$ of the cylinder,
\begin{equation}\label{Eq3}
\phi=\frac{x-x_0}{R},
\end{equation}
where $x_0$ specifies the choice of origin along the circumference. The embedding of the rolled geometry into three-dimensional Cartesian coordinates $(x',y',z')$ is then
\begin{equation}\label{Eq4}
x'=R\cos\phi,\qquad
y'=R\sin\phi,\qquad
z'=y.
\end{equation}
Equivalently, the inverse mapping is
\begin{equation}\label{Eq5}
x=x_0+R\phi,\qquad y=z'.
\end{equation}
Since the angular variable is defined modulo $2\pi$, the planar coordinate $x$ is defined modulo the tube circumference $2\pi R$.

We now consider a twofold rotation about the $x'$ axis,
\begin{equation}\label{Eq6}
[E\|C_{2,x'}]:\quad (x',y',z')\rightarrow (x',-y',-z').
\end{equation}
Using Eq. \eqref{Eq4}, this transformation implies
\begin{equation}\label{Eq7}
R\cos\phi \rightarrow R\cos\phi,\qquad
R\sin\phi \rightarrow -R\sin\phi,\qquad
z'=\rightarrow -z'.
\end{equation}
Therefore,
\begin{equation}\label{Eq8}
\phi\rightarrow-\phi \quad (\mathrm{mod}\ 2\pi),\qquad y\rightarrow-y.
\end{equation} 
Substituting into Eq. \eqref{Eq5}, we obtain
\begin{equation}\label{Eq9}
x\rightarrow2x_0-x,
\end{equation}
and hence
\begin{equation}\label{Eq10}
(x,y)\xrightarrow{\,[E\|C_{2,x'}]\,}(2x_0-x,-y).
\end{equation}

Similarly, for a twofold rotation about the $y'$ axis,
\begin{equation}\label{Eq11}
[E\|C_{2,y'}]:\quad (x',y',z')\rightarrow (-x',y',-z'),
\end{equation}
we have
\begin{equation}\label{Eq12}
R\cos\phi\rightarrow-R\cos\phi,\qquad
R\sin\phi\rightarrow R\sin\phi,\qquad
z'=-z'.
\end{equation}
A solution is
\begin{equation}\label{Eq13}
\phi\rightarrow\pi-\phi \quad (\mathrm{mod}\ 2\pi).
\end{equation}
Accordingly,
\begin{equation}\label{Eq14}
x\rightarrow2x_0+\pi R-x.
\end{equation}
Thus,
\begin{equation}\label{Eq15}
(x,y)\xrightarrow{\,[E\|C_{2,y'}]\,}(2x_0+\pi R-x,-y).
\end{equation}
Equations~\eqref{Eq10} and \eqref{Eq15} show that both $[E\|C_{2,x'}]$ and $[E\|C_{2,y'}]$ act as inversion of the planar coordinates up to an additive constant along the rolled direction. This constant depends only on the choice of origin $x_0$ and on the angular reference of the cylindrical parametrization. In particular, by choosing the planar origin such that the inversion center is located at $x=x_0$ for $[E\|C_{2,x'}]$, or at $x=x_0+\pi R/2$ for $[E\|C_{2,y'}]$, these transformations reduce to the standard planar inversion
\begin{equation}\label{Eq16}
(x,y)\rightarrow(-x,-y).
\end{equation}

Therefore, the twofold rotations $[E\|C_{2,x'}]$ and $[E\|C_{2,y'}]$ in the nanotube geometry are equivalent to inversion in the parent planar system, up to an origin shift along the circumference. Since such a shift only changes the choice of coordinate origin, the symmetry content is unchanged. Spin-space symmetries involving translation or inversion in the parent planar system therefore remain effective after rolling. Consequently, magnetic nanotubes inherit the symmetry properties of their parent phases and may be regarded as one-dimensional descendants of the corresponding two-dimensional magnets.

The symmetries and their consequences for the band structure are summarized in Table~\ref{table1} for collinear tubes.

\begin{table*}[htbp]
\centering
\caption{Critical symmetries and resulting band properties of coplanar magnetic nanotubes}
\label{table2}
\setlength{\tabcolsep}{8pt}
\begin{tabular}{c c c c}
\hline\hline
Parent magnetic order & Geometry & Symmetry & Band property \\
\hline\hline

FM & Achiral & $[C_{l,z'}\| C_{l,z'}]$,$[C_{2,z'}{\mathcal T}\| {\mathcal M}_{z'}{\mathcal T}]$, $[{\mathcal T}\|{\mathcal P}{\mathcal T}]$, $[{\mathcal T}C_{l,z'}\|{\mathcal P}{\mathcal T}C_{l,z'}]$ & $\omega({\bf s},{\bf k})=\omega({\bf s},-{\bf k})$, $\omega({\bf s},{\bf k})=\omega({\bf -s},{\bf k})$ \\
   & Chiral  & $[C_{2,z'}{\mathcal T}\|{\mathcal T}]$, $[C_{l,z'}\|C_{l,z'}| {\bf t}_{l}]$, $[C_{2,\bot}\|C_{2,\bot}]$ & $\omega({\bf s},{\bf k})\neq\omega({\bf s},-{\bf k})$, $\omega({\bf s},{\bf k})=\omega({\bf -s},{\bf k})$ \\

\hline\hline

AFM & Achiral & $[C_{l,z'}\| C_{l,z'}]$, $[C_{2,z'}{\mathcal T}\|{\mathcal T}]$, $[{\mathcal T}\|{\mathcal P}{\mathcal T}]$ & $\omega({\bf s},{\bf k})=\omega({\bf s},-{\bf k})$, $\omega({\bf s},{\bf k})=\omega({\bf -s},{\bf k})$ \\
    & Chiral  & $[C_{2,z'}{\mathcal T}\|{\mathcal T}]$, $[C_{l,z'}{\mathcal T}\|C_{l,z'}{\mathcal T}| {\bf t}_{l}]$, $[C_{l,z'}\|C_{l,z'}| {\bf t}_{l}]$, $[C_{2,\bot}\|C_{2,\bot}]$& $\omega({\bf s},{\bf k})\neq\omega({\bf s},-{\bf k})$, $\omega({\bf s},{\bf k})=\omega({\bf -s},{\bf k})$ \\

\hline\hline

AM & Achiral & $[C_{l,z'}\| C_{l,z'}]$, $[C_{2,z'}{\mathcal T}\|{\mathcal T}]$, $[{\mathcal T}C_{2l,z'}\| C_{2l,z'}{\mathcal P}{\mathcal T}]$ & $\omega({\bf s},{\bf k})=\omega({\bf s},-{\bf k})$, $\omega({\bf s},{\bf k})=\omega({\bf -s},{\bf k})$ \\
   & Chiral  & $[C_{2,z'}{\mathcal T}\|{\mathcal T}]$, $[C_{l,z'}\|C_{l,z'}| {\bf t}_{l}]$, $[C_{2,\bot}\|C_{2,\bot}]$ & $\omega({\bf s},{\bf k})\neq\omega({\bf s},-{\bf k})$, $\omega({\bf s},{\bf k})=\omega({\bf -s},{\bf k})$ \\

\hline\hline
\end{tabular}
\end{table*}

For noncollinear but coplanar nanotubes, the situation is different because the magnetic order is reconstructed by curvature. The screw symmetry remains valid only as a combined operation involving both real and spin spaces: the real-space screw operation must be accompanied by the corresponding spin-space rotation to leave the magnetic texture invariant. In contrast, the perpendicular twofold rotation $[E||C_{2,\perp}]$ is no longer a symmetry of the coplanar nanotube. Nevertheless, the combined spin-space and real-space rotation $[C_{2,\perp}||C_{2,\perp}]$ remains preserved. Since the nanotube supports only a single spin component $s_{z'}$ and a single momentum component $k_{z'}$, this symmetry maps $(s_{z'},k_{z'})\rightarrow(-s_{z'},-k_{z'})$, imposing a time-reversal-like constraint on the one-dimensional band structure.

The symmetries and their consequences for the band structure are summarized in Table~\ref{table2} for noncollinear tubes.

\textit{Appendix C: Effects of spin-orbit coupling---}
In this section, we investigate how relativistic spin-orbit coupling modifies the magnetic band structures of nanotubes. Specifically, we introduce a finite Rashba spin-orbit interaction into the tight-binding model,
\begin{equation}\label{Eq17}
\begin{aligned}
H_{sc}=\frac{i\lambda}{a} \sum_{i,j,\sigma,\sigma'}{\bf e}_{\lambda}\cdot({\bf \sigma} \times {\bf d}_{ij})_{\sigma\sigma'}c^{\dagger}_{i\sigma}c_{j\sigma'}.
\end{aligned}
\end{equation}
Here, ${\bf e}_{\lambda}$ denotes the Rashba axis, $\lambda$ is the Rashba spin-orbit coupling strength, and ${\bf d}_{ij}$ is the nearest-neighbor bond vector connecting sites $i$ and $j$. In nanotubes, ${\bf e}_{\lambda}$ points along the radial direction because curvature induces an intrinsic surface potential gradient along the surface normal. In general, spin-orbit coupling breaks the spin symmetries of the system, such that the magnetic properties must be described using magnetic groups rather than spin groups. Clearly, in achiral nanotubes, inversion symmetry is protected not only by the spin-group symmetry $[C_{2,\perp}{\mathcal T}||{\mathcal T}]$, but also present in the magnetic group (see Table \ref{table1}). Therefore, the inclusion of spin-orbit coupling in achiral ferromagnetic and altermagnetic nanotubes does not change our conclusions. In achiral antiferromagnetic nanotubes, the combined symmetry $[{\mathcal T}||{\mathcal P}{\mathcal T}]$ further protects Kramers degeneracy even in the presence of spin-orbit coupling. In contrast, for chiral nanotubes, the inversion symmetry in momentum space originates solely from the spin-group symmetries $[C_{2,\perp}{\mathcal T}||{\mathcal T}]$ and $[E||C_{2,\perp}]$. Since this symmetry is generally not preserved in the relativistic regime, the corresponding band structure can be modified by spin-orbit coupling, as discussed in Sec.~IV of the SM \cite{SM}. These SOC-induced modifications should be viewed as relativistic corrections to the nonrelativistic texture-induced mechanism emphasized in the main text.

We next consider the noncollinear but coplanar tubes. Similar to the collinear case, inversion symmetry remains allowed for achiral nanotubes in the relativistic regime. For chiral nanotubes, the time-reversal-like band-structure constraint is protected by both spin and magnetic symmetries, as shown in Table \ref{table2}. The nonrelativistic $p$-wave spin splitting therefore remains the dominating partial-wave in the presence of spin-orbit coupling. See Sec.~IV of the SM \cite{SM} for details.
\end{document}